\documentclass[12pt]{article}
\usepackage{a4}
\usepackage{hyperref}
\usepackage{amssymb}
\usepackage{cite}
\usepackage{color}
\usepackage{soul}

\sethlcolor{yellow}

\newcommand{\be}{\begin{equation}}
\newcommand{\ee}{\end{equation}}
\newcommand{\bea}{\begin{eqnarray}}
\newcommand{\eea}{\end{eqnarray}}

\begin{document}
\thispagestyle{empty}
\def\thefootnote{\fnsymbol{footnote}}
\begin{flushright}
{\sf  CERN-PH-TH/2013-013}\\

     \vskip 2em
\end{flushright}
\vskip 2.0em

\begin{center}\Large
Defects, Super-Poincar\'{e} line bundle and Fermionic T-duality. \\
\vskip 2em
January 2013
\end{center}\vskip 0.2cm
\begin{center}
Shmuel Elitzur$^1$,
\footnote{elitzur@vms.huji.ac.il
}
 Boaz Karni$^1$,\footnote{boazkarni@gmail.com
}
 Eliezer Rabinovici$^{1,3}$\footnote{eliezer@vms.huji.ac.il
}
and Gor Sarkissian$^{1,2}$\footnote{ gor.sarkissian@ysu.am
}
\end{center}
\vskip 0.2cm
\begin{center}
$^1$The Racah Institute of Physics,\  The Hebrew University,\\
Givat Ram, 91904, \ Jerusalem\\
Israel
\end{center}

\begin{center}
$^2$Department of Theoretical Physics, \ Yerevan State University,\\
Alex Manoogian 1, 0025\, Yerevan\\
Armenia
\end{center}
\begin{center}
$^3$ CERN, 1211 Geneva 23, Switzerland
\end{center}
\vskip 1.5em
\begin{abstract} \noindent
Topological defects are interfaces joining  two conformal field theories, for which the
 energy momentum tensor is continuous across the interface. A class of the topological defects is provided
by the interfaces separating two bulk systems each described by its own Lagrangian, where the two descriptions are related by a discrete symmetry.
  In this paper we elaborate on the cases in which the discrete symmetry is a bosonic or a fermionic T- duality. We review how the equations of motion imposed by the defect encode the general bosonic T- duality transformations for toroidal compactifications. We generalize this analysis in some detail to the case of topological defects allowed in coset CFTs, in particular to those cosets where the gauged group is either an axial or vector U(1). This is discussed in both the operator and Lagrangian approaches. We proceed to construct a defect encoding a fermionic T-duality. We show that the fermionic T-duality is implemented by the Super-Poincar\'{e} line bundle. The observation that the exponent of the gauge invariant flux on a defect is a kernel of the Fourier-Mukai transform of the Ramond-Ramond fields, is generalized to a fermionic T-duality.  This is done via a fiberwise integration on supermanifolds.\\
\end{abstract}

\newpage
\section{Introduction}

Interfaces in two-dimensional conformal field theories are playing a role in various topics, see e.g.
\cite{Oshikawa:1996dj,Petkova:2000ip,Petkova:2001zn,Petkova:2001ag,Bachas:2001vj,Fuchs:2002cm,Quella:2002ct,Graham:2003nc,Bachas:2004sy,
Frohlich:2006ch,Quella:2006de,Bachas:2007td,Fuchs:2007tx,Fuchs:2007fw,Runkel:2008gr,Brunner:2008fa,
Sarkissian:2008dq,Sarkissian:2009aa,Sarkissian:2009hy,Petkova:2009pe,Frohlich:2009gb,Drukker:2010jp,Fredenhagen:2010zh,Kapustin:2010zc,
Sarkissian:2010bw,Suszek:2011hg,Sarkissian:2011tr,Satoh:2011an,Bachas:2012bj,Fuchs:2012dt,Gaiotto:2012wh,Suszek:2012df,Carqueville:2012dk}.

Interfaces are oriented lines separating two different quantum filed theories.
In this paper we consider special class of interfaces, for which  the energy-momentum tensor is continuous across the defect.
Denoting the left- and right- moving energy-momentum tensors of the two theories by $T^1$, $T^2$, and $\bar{T}^1$, $\bar{T}^2$,
this condition takes the form:
\be
T^{(1)}=T^{(2)}\, ,\hspace{1cm} \bar{T}^{(1)}=\bar{T}^{(2)} \label{TopDef1}
\ee

Inserting a defect/interface  in the path integral is equivalent in the operator language to the insertion of an operator $D$ which maps the Hilbert space of CFT $1$ to that of CFT $2$. Thus  a defect can be described by such an operator.
  Condition (\ref{TopDef1}) should be considered
as  implying also that the corresponding operator $D$ commutes with the Virasoro modes:
\be\label{topcom}
D L^1_m=L^2_m D\hspace{1cm}{\rm and}\hspace{1cm} D \bar{L}^1_m=\bar{L}^2_m D
\ee

where $L^i_m$ and $\bar{L}^i_m$ act on the state space ${\cal H}_i$, $i=1,2$,
and therefore the interface  can be continuously deformed without affecting the value of correlators
as long it does not cross any field insertion point. These interfaces are called topological defects \cite{Bachas:2004sy}.
Topological defects have the following properties.
\begin{itemize}
\item
Two topological defects can be moved and merged with each other to create a new defect. In the operator language
 the defect fusion corresponds to the composition  of the defect operators \cite{Petkova:2000ip,Petkova:2001zn,Petkova:2001ag,Frohlich:2006ch}.
\item
Similarly a topological defect can be moved to the boundary and  fused
with it, producing new boundary conditions\cite{Petkova:2001zn,Graham:2003nc,Quella:2006de}.
The new boundary state is given by the action of the defect operator on the boundary state.
Remembering that in String theory boundary states correspond
to D-branes, one arrives to the conclusion that topological defects induce  D-brane transformations.
On the other hand D-branes are classified by their Ramond-Ramond or K-theory charges.
Therefore topological defects should induce also transformations in cohomology or K-theory groups.
It is expected that this transform should be of the Fourier-Mukai type \cite{Fuchs:2007fw,Brunner:2008fa,Sarkissian:2008dq,Sarkissian:2010bw,Bachas:2012bj}.

\end{itemize}

Let us now take a closer look at the equations (\ref{TopDef1}) and (\ref{topcom}).

As follows from the above discussion $D$ maps  an eigenstate $|a\rangle$ of the $L_0^1$
to the eigenstate of $L_0^2$, with the same eigenvalue, if $|a\rangle$ is not in the null space of $D$. Hence the left and right  Hamiltonians of the two theories coincide on the pair of states $(|a\rangle,D|a\rangle)$, where
$|a\rangle$ is an eigenstate of the left or  right Hamiltonian of the first theory belonging to ${\cal H}_1/{\rm Null}(D)$. Thus
the theories admitting a topological defect to join them,
 have the same spectra once restricted to the subspace $ \{(|a\rangle,D|a\rangle)\,\,\,  |\,\,\, |a\rangle \in {\cal H}_1 /{\rm Null}(D)\}$.

In this paper we will analyze the defects also in the Lagrangian formalism. In the Lagrangian approach to defects, one has besides the bulk equation of motion,
also defect equations of motion \cite{Runkel:2008gr,Sarkissian:2008dq}. The defect equations of motion schematically have the form
$F(\Phi_1,\Phi_2,\Psi)=0$, where $F$ is some function, $\Phi_i$ is a collective notation for the fields of the first and the second theories, and $\Psi$ is a collective notation for fields on the defect that are not inherited from the bulk.
The energy-momentum tensors coincide when the defect equations of motion are imposed.
Comparing this with the operatorial picture, we see that the defect equations of motion capture the information on the structure of the defect.

Presently we are not aware of a complete classification of theories that can be joined by a topological defect, aside from the necessary condition that the theories must have the same central charges.
We would like, however, to mention some typical situations where that is possible.

The defects with a trivial null space describe theories with the same spectra.
Hence these defects exploit and uncover various symmetries
of the theory.
In particular such defects
connect different duality pictures of the same CFT, like those related by  T-duality \cite{Giveon:1994fu} and mirror symmetry \cite{Bachas:2007td,Fuchs:2007tx,Sarkissian:2008dq,Kapustin:2010zc,Bachas:2012bj}.
Let us note the following properties of these defects.
As explained above the defect relates eigenstates of the Hamiltonians with the
coinciding values of the Hamiltonians. Therefore the defect equations of motion
should produce the corresponding duality relations \cite{Sarkissian:2008dq,Kapustin:2010zc}.
We demonstrate this point in this paper
in some new  instances.

 There are several examples of theories connected by defects with a non-trivial null space. Among them are
\begin{itemize}
\item
Scalars compactified on  circles at the rationally related radii \cite{Fuchs:2007tx,Bachas:2007td}.
\item
Torus compactifications related by $O(d,d|\mathbb{Q})$
 semi-group transformations with rational entries \cite{Bachas:2012bj}.
\item
Theories  related by orbifold constructions \cite{Frohlich:2006ch,Bachas:2007td,Sarkissian:2008dq}.
\item
different modular invariants constructed out of the same chiral data
  \cite{Frohlich:2009gb,Carqueville:2012dk}.
\end{itemize}
In the case of defects in rational conformal field theory a relation between the corresponding theories  in the terms of the underlying modular tensor categories was suggested in \cite{Fuchs:2012dt}.

In this paper we
\begin{itemize}
\item
Discuss  topological defects joining axial and vector gaugings of $G/U(1)$ gauged WZW models.
\item
Generalize the construction of defects joining theories related by bosonic T-duality to the case of superspace target spaces related by a fermionic T-duality.
\end{itemize}

The paper is organized in the following way.

In section 2 we review topological defects implementing bosonic T-duality \cite{Giveon:1994fu}.

In section 3 we construct defects  between axial and vector gauging of $G/U(1)$ gauged WZW models \cite{Altschuler:1987zb,Bardakci:1987ee,Gawedzki:1988nj,Nahm:1988sn} for a general group $G$. For the case of $G=SU(2)$ \cite{Bardacki:1990wj} the geometrical construction is translated to the algebraic parafermionic language.
We show that for level $k$ parafermions there are $k+1$ topological defects mapping axially gauged $SU(2)/U(1)$ cosets
to the vectorially gauged $SU(2)/U(1)$ coset, labeled by the integrable spin $j=0,\ldots,{k\over 2}$. We construct them in both the Lagrangian approach,
and  algebraic one, in the latter by identifying the appropriate operators in the parafermion Hilbert space.
We show that the defect corresponding to $j=0$  implements $\mathbb{Z}_k$ orbifolding together with  T-duality. These defects project $A_{j,n}$ Cardy branes in $SU(2)/U(1)$ coset
to the $B_j$ branes constructed in \cite{Maldacena:2001ky}.

In section 4 we  study the defect performing the fermionic T-duality \cite{Berkovits:2008ic}.
It is established that the defect implementing bosonic T-duality is given by the Poincar\'{e} line bundle \cite{Fuchs:2007fw,Sarkissian:2008dq}.
We show that the defect inducing the fermionic T-duality  is given by the fermionic
generalization of the Poincar\'{e} line bundle, which we denote as  Super-Poincar\'{e} line bundle.
We demonstrate that the defect equations of motion reproduce the fermionic T- duality transformation rules  found in \cite{Berkovits:2008ic}.
Using the exponent of the gauge invariant flux on this defect as a kernel of the Fourier-Mukai transform
with a pushforward map given by the fiberwise integration on supermanifold, we derive the transformation of the
Ramond-Ramond fields under the fermionic T-duality.

In four appendices  A, B, C and D some calculations and constructions are explained in more detail.

\section{Topological defects and bosonic T-duality}
In this section we review some basic facts concerning topological defects and their relation to  T-duality.
We first use the definition of the topological defects  in the simple example of a scalar field compactified on a circle to demonstrate how the defect equations of motion together with the requirement to be topological
reproduce the appropriate duality transformations.
In the next subsection we generalize this to the factorized T-duality in non-linear sigma models with  isometries.
In these cases the null space of the defects is trivial and the defects are invertible. We then go on to discuss cases where the null space is non trivial.
We present a defect generating a combined action of the  $\mathbb{Z}_k$ orbifolding together with a T-duality transformation.
Then we review defects implementing generators
of the full $O(d,d|\mathbb{Z})$ duality group in the case of toroidal compactification.
These defects are invertible as well.
We conclude this section explaining how
the T-duality transformation of the Ramond-Ramond charges can be written as the Fourier-Mukai transform
with the kernel given by the exponent of the gauge invariant
flux on the corresponding topological defect.

\subsection{Preliminaries}
Defects in two-dimensional quantum field theory are oriented lines
separating different quantum field theories, labeled (in this paper) by 1 and 2.
Conformal defects are required to satisfy \cite{Bachas:2001vj}

\be
T^{(1)}-\bar{T}^{(1)}=T^{(2)}-\bar{T}^{(2)} \label{ConfDef}
\ee

Topological defects satisfy \cite{Petkova:2000ip}
\be
T^{(1)}=T^{(2)}\, ,\hspace{1cm} \bar{T}^{(1)}=\bar{T}^{(2)} \label{TopDef}
\ee
Since the stress-energy tensor is a generator of diffeomorphisms, condition (\ref{TopDef})
implies that the defect is invariant under a distortion of the line to which it is
attached. A notion of fusion between a defect and a boundary can be expected in the case
of topological defects, since the latter can be moved to the boundary without changing the correlator \cite{Petkova:2001zn}.

We review the construction of an action with defects \cite{Fuchs:2007fw,Sarkissian:2008dq}. We locate the defect at the vertical line S defined by the condition $\sigma=0$.
Denote by $\Sigma_1$ the left half-plane $(\sigma\leq 0)$, and by $\Sigma_2$ the right half-plane $(\sigma\geq 0)$, and
a pair of maps $X: \Sigma_1\rightarrow M_1$ and $\tilde{X}: \Sigma_2   \rightarrow M_2$, where $M_1$ and $M_2$ are the target spaces for the two quantum field theories.
Suppose we have a submanifold $Q$ of the cartesian product of target spaces: $Q\subset M_1\times M_2$, with a
connection one-form $A$, and a combined map :
\bea
\Phi : S\rightarrow M_1\times M_2\\ \nonumber
s\mapsto (X(s), \tilde{X}(s))
\eea
which takes values  in the submanifold $Q$. $Q$ is called the world-volume of the defect.

 In this setup we can write the action:
\be\label{defact}
I=\int_{\Sigma_1}dx^+dx^-L_1+\int_{\Sigma_2}dx^+dx^-L_2+\int_S \Phi^*A
\ee
where
\be
L_1=E^{(1)}_{mn}\partial X^m\bar{\partial}X^n,
\ee

\be
L_2=E^{(2)}_{mn}\partial \tilde{X}^m\bar{\partial}\tilde{X}^n,
\ee

\be\label{lccor}
x^{\pm}=\tau\pm \sigma,
\ee
with $E^{(i)}_{mn}$ being the components of two  second rank tensors. The tensors $E^{(i)}$ are split as
\be
E^{(i)}=G^{(i)}+B^{(i)}.
\ee
where $G^{(i)}$ are the symmetric target space metrics of the two sigma models and $B^{(i)}$ are the corresponding NS antisymmetric two-forms.

As a warm-up exercise we work out the following simple example, when we have on both sides free scalars compactified on circles  $S^1_X$ and $S^1_{\tilde{X}}$ of radii $R_1$ and $R_2$:
\be
L_1=R_1^2\partial X\bar{\partial}X
\ee
and
\be
L_2=R_2^2\partial \tilde{X}\bar{\partial}\tilde{X}.
\ee
The world-volume of the defect is a product of the target spaces $S^1_X\times S^1_{\tilde{X}}$ with the connection  $A=-Xd\tilde{X}$. The curvature of this connection is $F=d\tilde{X}\wedge dX$. This forms a Poincar\'{e} bundle ${\cal P}$ \cite{Griffith}.
 The equations of motion on the defect line are:
\be\label{r1}
R_1^2(\partial X-\bar{\partial} X)-\partial_{\tau}\tilde{X}=0
\ee
\be\label{r2}
R_2^2(\partial \tilde{X}-\bar{\partial} \tilde{X})-\partial_{\tau}X=0
\ee
For $R_2={1\over R_1}$,  (\ref{r1}) and (\ref{r2}) take the form:
\be\label{xr1}
R_1^2(\partial X-\bar{\partial} X)-(\partial \tilde{X}+\bar{\partial} \tilde{X})=0
\ee
\be\label{yr2}
(\partial \tilde{X}-\bar{\partial} \tilde{X})-R_1^2(\partial X+\bar{\partial} X)=0
\ee
Equations (\ref{xr1}) and (\ref{yr2}) imply
\be\label{tx}
R_1^2\partial X=\partial \tilde{X}
\ee
\be\label{ty}
R_1^2\bar{\partial} X=-\bar{\partial} \tilde{X}
\ee
which are the  T-duality relations.  Equations (\ref{tx}) and (\ref{ty})  also show that the defect given by the Poincar\'{e} bundle ${\cal P}$
for $R_2={1\over R_1}$ is topological. If this is not the case, then from equations (\ref{r1}) and (\ref{r2}) one can derive equation (\ref{ConfDef}) and the defect is conformal, but not topological.

One generalization that comes to mind is a defect ${\cal P}^k$ with the same world-volume but with $k$ units of the flux above:
$F=kd\tilde{X}\wedge dX$. In the same way it is possible to show that this defect is topological
when the radii satisfy the relation
\be\label{r1r2}
R_1R_2=k
\ee
and instead of (\ref{tx}) and (\ref{ty}) one obtains:
\be\label{txk}
R_1^2\partial X=k\partial \tilde{X}
\ee
\be\label{tyk}
R_1^2\bar{\partial} X=-k\bar{\partial} \tilde{X}
\ee

These relations imply that the defect ${\cal P}^k$  combines the actions of the $Z_k$ orbifolding
and T-duality.

All  this is in agreement with  \cite{Bachas:2007td,Fuchs:2007tx}, where more general submanifolds $Q$ are considered. There the worldvolume $Q$ of the defect is either two dimensional with flux $F=k_{1}d\tilde{X}\wedge dX$,  but allowed to wrap the product
$S^1_X\times S^1_{\tilde{X}}$ torus $k_2$ times, or $Q$ is made one dimensional winding around the cycles $(k_1,k_2)$ times. Then  the existence of the  topological defect is proved for the radii satisfying
the relations:
\be
R_1R_2=\left|{k_1\over k_2}\right|\hspace{1cm}{\rm or} \hspace{1cm}{R_2\over R_1}=\left|{k_1\over k_2}\right|
\ee
where $k_1,k_2\in \mathbb{Z}$.

\subsection{Factorized T-duality in non-linear sigma model}

Let us turn to the defect description of the T-duality arising when one has action
\be\label{actbg}
I=\int_{\Sigma}dx^+dx^-E_{mn}\partial X^m\bar{\partial}X^n,
\ee
on a  target space with the isometry  \cite{Buscher:1987qj,Giveon:1994fu}. Here, and in the following, repeated indices are summed over.
Suppose that the coordinate $X^1$ is chosen in the direction of the isometry. This means that  $G_{ij}$ and $B_{ij}$ do not depend on $X^1$.
It is known that in this situation the action with the background matrix $E$ is equivalent  to the action with the background matrix $\tilde{E}$,
where
\bea\label{tdual}
&&\tilde{E}_{11}={1\over E_{11}}\\ \nonumber
&&\tilde{E}_{1M}={E_{1M}\over E_{11}}\\ \nonumber
&&\tilde{E}_{M1}=-{E_{M1}\over E_{11}}\\ \nonumber
&&\tilde{E}_{MN}=E_{MN}-{E_{M1}E_{1N}\over E_{11}}
\eea
In components one has:
\bea\label{tdualc}
&&\tilde{G}_{11}={1\over G_{11}}\\ \nonumber
&&\tilde{G}_{1M}={B_{1M}\over G_{11}}\\ \nonumber
&&\tilde{B}_{1M}={G_{1M}\over G_{11}}\\ \nonumber
&&\tilde{G}_{MN}=G_{MN}-{1\over G_{11}} (G_{M1}G_{1N}+B_{1N}B_{M1})\\ \nonumber
&&\tilde{B}_{MN}=B_{MN}-{1\over G_{11}} (G_{M1}B_{1N}+G_{1N}B_{M1})
\eea
The capital latin indices run from $2$ to ${\rm dim} M$.

The dual coordinate $\tilde{X}^1$ is related to the original $X^1$ by the relations:
\be\label{xtilx}
\partial \tilde{X}^1=E_{11}\partial X^1 +E_{M1}\partial X^M\hspace{0.5cm}{\rm  and}\hspace{0.5cm}
\bar{\partial}\tilde{X}^1=-(E_{11}\bar{\partial}X^1+E_{1M}\bar{\partial} X^M)
\ee
The rest of the coordinates remains unchanged.

Consider the action  (\ref{defact}) with a defect as in the situation above, where $M$ and $\tilde{M}$ are related by the equations (\ref{tdual}),
$Q$ is the correspondence space, given by the equations
\be\label{xnyn}
X^N=\tilde{X}^N,\hspace{1cm} N=2\ldots {\rm dim} M
\ee
with the connection
\be\label{pbcon}
 A=-X^1d\tilde{X}^1
\ee
 and  the curvature
\be
 F=d\tilde{X}^1\wedge dX^1.
\ee
In this case the action (\ref{defact}) yields
\be\label{td1}
E_{j1}\partial X^j-E_{1j}\bar{\partial} X^j-\partial_{\tau}\tilde{X}^1=0
\ee

\be\label{td2}
E_{jN}\partial X^j-E_{Nj}\bar{\partial} X^j-\tilde{E}_{jN}\partial \tilde{X}^j+\tilde{E}_{Nj}\bar{\partial} \tilde{X}^j=0,\hspace{1cm} N=2\ldots {\rm dim} M
\ee

\be\label{td3}
\tilde{E}_{j1}\partial \tilde{X}^j-\tilde{E}_{1j}\bar{\partial} \tilde{X}^j-\partial_{\tau}X^1=0.
\ee
 The index $j$ runs from 1 to ${\rm dim} M$.
Additionally the conditions (\ref{xnyn}) imply
\be
\partial_{\tau}X^N=\partial_{\tau}\tilde{X}^N,\hspace{1cm} N=2\ldots {\rm dim} M
\ee
or in the coordinates (\ref{lccor}):
\be\label{dxt}
\partial X^N+\bar{\partial} X^N=\partial \tilde{X}^N+\bar{\partial} \tilde{X}^N,\hspace{1cm} N=2\ldots {\rm dim} M
\ee

Solving the equations  (\ref{td1}), (\ref{td2}), (\ref{td3}) and (\ref{dxt})  one obtains:
\bea\label{fineqt}
&&\bar{\partial} \tilde{X}^N=\bar{\partial} X^N\hspace{1cm} N=2,\ldots {\rm dim}M\\ \nonumber
&&\partial \tilde{X}^N=\partial X^N\hspace{1cm} N=2,\ldots {\rm dim}M\\ \nonumber
&&\partial \tilde{X}^1=E_{11}\partial X^1+E_{M1}\partial X^M\\ \nonumber
&&\bar{\partial} \tilde{X}^1=-(E_{11}\bar{\partial} X^1+E_{1M}\bar{\partial} X^M)
\eea
The details of calculations appear in appendix A. We see that equations  (\ref{fineqt})  coincide with the T-duality relations (\ref{xtilx}).
Therefore the defect given by the Poincar\'{e} bundle on the correspondence space induces T-duality.

One can check that (\ref{tdualc}) and (\ref{fineqt}) imply
\be
T=G_{ij}\partial  X^i\partial  X^j=\tilde{T}=\tilde{G}_{ij}\partial \tilde{X}^i\partial \tilde{X}^j
\ee
and
\be
\bar{T}=G_{ij}\bar{\partial} X^i\bar{\partial} X^j=\tilde{\bar{T}}=\tilde{G}_{ij}\bar{\partial} \tilde{X}^i\bar{\partial} \tilde{X}^j
\ee
which means that the defect is topological.

In this general set-up one can also consider the defect with the same world-volume
given by equations (\ref{xnyn}) but with the flux
\be\label{kflu}
F=kd\tilde{X}^1\wedge dX^1.
\ee
Repeating the calculations above one can show that this defect is topological if
$E$ and $\tilde{E}$ are related by the equations
\bea\label{tduall}
&&\tilde{E}_{11}={k^2\over E_{11}}\\ \nonumber
&&\tilde{E}_{1M}={kE_{1M}\over E_{11}}\\ \nonumber
&&\tilde{E}_{M1}=-{kE_{M1}\over E_{11}}\\ \nonumber
&&\tilde{E}_{MN}=E_{MN}-{E_{M1}E_{1N}\over E_{11}}
\eea
Again the effects of the $Z_k$ orbifolding of the first coordinate and the T-duality are combined.

All this can be   generalized to  T-dualizing of  several coordinates.
Suppose we T-dualize the first $n$ coordinates, indexed by Greek letters.
The matrix $E$ is  broken to four pieces:
\be
E=\left(\begin{array}{ccc}
E_{\alpha\beta}& E_{\alpha N}\\
E_{M\beta}& E_{MN}\end{array}\right)
\ee
The transformed background has the form
\be\label{trgen}
\tilde{E}=\left(\begin{array}{ccc}
E^{-1}_{\alpha\beta}& E^{-1}_{\alpha\beta}E_{\beta N}\\
-E_{M\alpha}E^{-1}_{\alpha\beta}&E_{MN}-E_{M\alpha} E^{-1}_{\alpha\beta}E_{\beta N}\end{array}\right)
\ee

Now we should consider the defect, with the world-volume
\be
X^N=\tilde{X}^N, \hspace{1cm} N=n+1,\ldots {\rm dim} M,
\ee
 with the connection
\be
A=-\sum_1^nX^{\alpha}d\tilde{X}^{\alpha}
\ee
 and the curvature
\be
F=\sum_1^nd\tilde{X}^{\alpha}\wedge d X^{\alpha}.
\ee
In the same way it can be shown that for $M$ and $\tilde{M}$ related by equations (\ref{trgen}) this defect is topological and implies the defect equations:
\bea
&&\bar{\partial} \tilde{X}^N=\bar{\partial} X^N \hspace{1cm} N=n+1,\ldots {\rm dim}M\\ \nonumber
&&\partial \tilde{X}^N=\partial X^N \hspace{1cm} N=n+1,\ldots {\rm dim}M\\ \nonumber
&&\partial \tilde{X}^{\alpha}=E_{\beta\alpha}\partial X^{\beta}+E_{M\alpha}\partial X^M\\ \nonumber
&&\bar{\partial} \tilde{X}^{\alpha}=-(E_{\alpha\beta}\bar{\partial} X^{\beta}+E_{\alpha M}\bar{\partial} X^M)
\eea
We have  obtained  again T-duality relations for  several T-dualized coordinates.

\subsection{Dualities of toroidal compactifications}
Dualities of the toroidal compactification form the $O(n,n,\mathbb{Z})$ group.
The generators of this group are  factorized dualities, integer shifts of the flux of $B$ fields and the integer basis changes \cite{Elitzur:1994ri, Giveon:1988tt, Giveon:1994fu}.
Defects inducing factorized dualities were discussed in the previous subsection.
 For completeness let us mention  defects inducing the $B$-flux  shift  and the integer basis change symmetries.

Consider diagonal defect
\be
X^i=\tilde{X}^i\hspace{1cm} i=1,\ldots , {\rm dim}M
\ee
 with flux $F$.
In this case equations of motion for the defect take the form:
\be\label{td2n}
E_{ji}\partial X^j-E_{ij}\bar{\partial} X^j-\tilde{E}_{ji}\partial \tilde{X}^j+\tilde{E}_{ij}\bar{\partial} \tilde{X}^j+F_{ij}\partial_{\tau}X^j=0
\ee
and additionally
\be\label{dxtn}
\partial X^i+\bar{\partial} X^i=\partial \tilde{X}^i+\bar{\partial} \tilde{X}^i
\ee
It can be seen that if the matrices $E$ and $\tilde{E}$ differ only in the $B$ field and the difference
is equal to $F$:
\bea
&&\tilde{G}=G\\ \nonumber
&&\tilde{B}=B-F\\
\eea
this defect is topological and implies:
\bea
&&\bar{\partial} \tilde{X}^i=\bar{\partial} X^i \\ \nonumber
&&\partial \tilde{X}^i=\partial X^i \hspace{1cm}
\eea
This example was considered in \cite{Sarkissian:2008dq}.

Another interesting example is given by a defect with world-volume  given by a linear
embedding:
\be
X^i=A^i_k\tilde{X}^k
\ee
and with no flux.
The defect equations of motion are:
\be\label{td2na}
(E^i_j\partial X^j-E^i_j\bar{\partial} X^j)A^i_k-\tilde{E}_{jk}\partial \tilde{X}^j+\tilde{E}_{kj}\bar{\partial} \tilde{X}^j=0
\ee
and additionally
\be\label{dxtna}
\partial X^i+\bar{\partial} X^i=A^i_k(\partial \tilde{X}^k+\bar{\partial} \tilde{X}^k)
\ee
One can verify that if $E$ and $\tilde{E}$ satisfy the relation:
\be
\tilde{E}_{mk}=E_{ji}A^i_k A^j_m
\ee
this defect is topological and the defect equations of motion are solved by
\bea
\partial X^i=A^i_k\partial \tilde{X}^k\\ \nonumber
\bar{\partial} X^i=A^i_k\bar{\partial} \tilde{X}^k
\eea
For the torus compactifications the Dirac's quantization condition of the flux $F_{ij}$ and the quantization  imposed on the matrix $A^i_k$
by the compactness of the defect
 bring to the appropriate  integer $B$-flux
shifts and the integer  basis change transformations \cite{Elitzur:1994ri, Giveon:1988tt, Giveon:1994fu}.
Considering multiply wrapped  defects leads to the extended semi-group $O(d,d,\mathbb{Q})$  of the defects  \cite{Bachas:2007td,Bachas:2012bj}


\subsection{Defects and Fourier-Mukai transform}
 As explained at the beginning, a topological defect can be fused with a boundary, producing a new  boundary condition from the old one.
On the other hand boundary conditions correspond to D-branes, which can be characterized by their Ramond-Ramond charges or more precisely
by an element of the K-theory. Therefore  an action of the defect  on the RR charges and K-theory elements should be defined.
The form of this action on RR charges turns out to be connected to the flux on the corresponding defect.
Mathematically this flux serves as a kernel of an operation known as Fourier-Mukai transformation \cite{huy,bbr}.

Consider for example the T-duality transformation of the Ramond-Ramond fields.

It is found in \cite{Hori:1999me} that the   T-duality transformation of the Ramond-Ramond fields RR fields \cite{Bergshoeff:1995as,Eyras:1998hn}
of the theory on $T^n\times M$ and those of the T-dual theory on $\hat{T}^n\times M$ are related by a Fourier-Mukai transform:

\be\label{fmt}
\hat{\cal G}=\int_{T^n}e^{\hat{B}-B+\sum_{i=1}^n d\hat{t}_i\wedge dt^i}{\cal G}
\ee
Here $B$ is the Neveu-Schwarz $B$-field and ${\cal G}=\sum_p {\cal G}_{p+2}$ is the sum of gauge invariant RR field strength where the sum is over
$p=0,2,4,\ldots$ for Type IIA and $p=-1,1,3,\ldots$ for Type IIB.
The integrand in (\ref{fmt}) is considered as a form on the space $M\times T^n\times \hat{T}^n$ and the
 fiberwise integration $\int_{T^n}$, maps  forms on  $M\times T^n\times \hat{T}^n$ to forms  on $M\times \hat{T}^n$. The integral operates on the forms of the highest degree  $n$ in $dt_i$ and sets to zero forms of lower degree in $dt_i$ \cite{bott}:
\bea\label{fbs}
f(x,\hat{t}_i,t^i)p^*\omega \wedge dt_{i_1}\wedge \ldots dt_{i_r}\mapsto 0, \hspace{1cm} r<n\\ \nonumber
f(x,\hat{t}_i,t^i)p^*\omega \wedge dt_{1}\wedge \ldots dt_{n}\mapsto \omega\int_{T^n}   f(x,\hat{t}_i,t^i)dt_{1}\ldots dt_{n}
\eea
Here $p$ is the projection $M\times T^n\times \hat{T}^n\to M\times \hat{T}^n$,  $\omega$ is a form on $M\times \hat{T}^n$, $f(x,\hat{t}_i,t^i)$
is an arbitrary function
and $x$ denotes a point in $M$. The fiberwise integration (\ref{fbs}) is actually the Berezin integration, which is not surprising
when one remembers that the one-forms $dt_i$ anticommute.

Note  that the kernel of the Fourier-Mukai transform (\ref{fmt}) is indeed  the exponent of the gauge invariant combination of the $B$ fields and the flux of the T-duality defect
\be
e^{\cal F}=e^{\hat{B}-B+\sum_{i=1}^n d\hat{t}_i\wedge dt^i}
\ee

Let us check that in the simple case of T-dualizing of  one coordinate and without a  $B$ field, that formula (\ref{fmt})
yields the known map of $Dp$ to $D(p\pm 1)$ branes.
In this case Eq. (\ref{fmt}) takes the form
\be\label{gtt}
\hat{\cal G}=\int_{S^1} {\cal G} e^{dt\wedge d\hat{t}}=\int_{S^1} {\cal G}(1+dt\wedge d\hat{t})
\ee
Suppose that the $Dp$-brane is transverse to the coordinate $t$ and therefore the volume- form ${\cal G}$ does not contain $dt$. In this case
(\ref{gtt}) according to (\ref{fbs})  implies
\be
\hat{\cal G}={\cal G}\wedge d\hat{t}
\ee
and thus $\hat{\cal G}$ describes $D(p+1)$-brane as expected. Now consider the case when $Dp$-brane contains the coordinate $t$
and therefore the volume-form ${\cal G}$ has the form ${\cal G}=\omega\wedge dt$. In this case  Eq. (\ref{gtt}) yields
\be
\hat{\cal G}=\omega
\ee
and represents $D(p-1)$-branes again in agreement with   T-duality.

\section{Defects between vectorially and axially gauged WZW models}

In this  section we construct topological defects mapping the axially gauged ${G\over U(1)}_{\rm axial}$  WZW model to the
vectorially gauged ${G\over U(1)}_{\rm vectorial}$ WZW model for a general group $G$. For the case $G=SU(2)$ we analyze  the corresponding operators
acting in the Hilbert space of parafermions and find that for the level $k$  parafermions there are  $k+1$
such topological defects, labeled by the integrable spin $j=0,\ldots ,{k\over 2}$.
This is another example of the case of a non trivial null space for the defect.
The object is to realize these defects
in the Lagrangian approach as a line separating axially and vectorially gauged WZW models.
This problem is solved in this section. First we present the geometrical ansatz for the defects (formula (\ref{qman})
below) and check that it leads to the  action that glues axially  and vectorially gauged models.
Then we study in detail the defect given by $j=0$ and show that it coincides with the defect
 with the flux (\ref{kflu}), studied in the previous section, and implements $\mathbb{Z}_k$ orbifolding together with the T-duality.
In the rest of the section we construct defects as operators in the Hilbert space of the parafermions.
In appendix B, we calculate the overlap of these operators with the eigen-position state and show that they have the
geometry of the ansatz (\ref{qman}).

\subsection{Geometry and flux of the  defects  gluing axially-vectorially  gauged models}

The action of the gauged WZW model is \cite{Altschuler:1987zb,Bardakci:1987ee,Gawedzki:1988nj,Nahm:1988sn}:
\be\label{gauact}
S^{G/H}(g,A)=S^{\rm WZW}+S^{\rm gauge}\, ,
\ee
where
\bea
S^{\rm WZW}(g)&=&{k \over 4\pi}\int_{\Sigma}{\rm Tr}(\partial_{+} g\partial_{-}g^{-1})dx^+dx^-
 +{k \over 4\pi}\int_B {1\over 3}{\rm tr}(g^{-1}d g)^3\\ \nonumber
&\equiv& { k\over{4 \pi}} \left[ \int_{\Sigma}dx^+dx^- L^{\rm kin}
+ \int_B \omega^{\rm WZ}\right] \, ,
 \eea
\be
S^{\rm gauge}={k\over 2\pi }\int_{\Sigma}L_v^{\rm gauge}dx^+dx^-\, ,
\ee
\be
L_v^{\rm gauge}(g,A)={\rm tr}[-g^{-1}\partial_{+}gA_{-}+\partial_{-}gg^{-1}A_{+}+A_{-}g^{-1}A_{+}g-A_{+}A_{-}]\, .
\ee
Here $H$ is subgroup of $G$, $g\in G$ and $B$ is a 3-manifold such that $\partial B=\Sigma$ and $A$ is a gauge field taking values in the $H$ Lie algebra. \\
Using  the Polyakov-Wiegmann identities:
\be
\label{pwk}
L^{\rm kin}(g h) =  L^{\rm kin}(g) + L^{\rm kin} (h)-
 \left(  {\rm Tr} \big(g^{-1}\partial_+ g \partial_{-} h h^{-1}\big)+
 {\rm Tr} \big(g^{-1} \partial_{-} g\partial_+  h h^{-1}\big)\right)\, ,
\ee
\be
\label{pwwz}
\omega^{\rm WZ}(g h) = \omega^{\rm WZ}(g) + \omega^{\rm WZ}(h)
 - {\rm d}\Big({\rm Tr} \big(g^{-1} {\rm d}g  {\rm d}h h^{-1}\big)\Big)\, ,
\ee

it is possible to verify that the action (\ref{gauact}) is invariant under the gauge transformation:
\be
g\rightarrow hgh^{-1}\, ,\hspace{1cm} A\rightarrow hAh^{-1}+dhh^{-1}
\ee
for $h: \Sigma\rightarrow H$.
This is a vectorially gauged model.

For the case of $H=U(1)$ considered here
there exists the system is axially gauge invariant under the transformations
\be
g\rightarrow hgh\, ,\hspace{1cm} A\rightarrow A+dhh^{-1}
\ee
for $h: \Sigma\rightarrow U(1)$.
In the axially gauged model the gauge field dependent term is
\be
L_a^{\rm gauge}(g,A)={\rm tr}[g^{-1}\partial_{+}gA_{-}+\partial_{-}gg^{-1}A_{+}-A_{-}g^{-1}A_{+}g-A_{+}A_{-}]\, .
\ee
There are several steps needed in order to write a well defined action on the defect, with an image in the submanifold $Q\subseteq G\times G$
\be\label{defcond}
S\rightarrow G\times G: s\mapsto (g_1(s), g_2(s))\in Q,
\ee
with a defect line $S$ separating  vectorially and axially gauged models, in the presence of a WZW form \cite{Fuchs:2007fw}.\\
First, there should exist a two-form $\varpi$ satisfying the relation
\be\label{varka}
d\varpi(g_1,g_2)=\omega^{\rm WZ}(g_1)|_Q-\omega^{\rm WZ}(g_2)|_Q
\ee
Second, one should introduce an auxiliary disc $D$ satisfying the conditions:
\be
\partial B_1=\Sigma_1 \cup D\hspace{0.5cm}{\rm and} \hspace{0.5cm}\partial B_2=\Sigma_2 \cup \bar{D},
\ee
where the unions are such that $\partial \Sigma_1 = \partial D = S$ and $\partial \Sigma_2 = \partial \bar{D} = \bar{S}$, but the orientations of the gluing are opposite.

The fields $g_1$ and $g_2$ are extended to this disc while holding the condition
(\ref{defcond}). After this preparations the topological part of the action takes the form\cite{Fuchs:2007fw}
\be
S^{\rm top-def}={k \over 4\pi}\int_{B_1} \omega^{\rm WZ}(g_1)+
{k \over 4\pi}\int_{B_2} \omega^{\rm WZ}(g_2)-{k \over 4\pi}\int_D\varpi(g_1,g_2)
\ee
One should choose an appropriate $Q$. One of the requirements is that $Q$ would be invariant
under the vector and axial transformations.
We suggest the following ansatz:
\be\label{qman}
(g_1, g_2)=(C_{\mu}p,L_1pL_2)
\ee
Here $p\in G$, $L_1\in U(1)$, $L_2\in U(1)$ and $C_{\mu}$ is a conjugacy class
\be
C_{\mu}=le^{2i\pi\mu/k}l^{-1},\hspace{0.2cm} l\in G
\ee
where $\mu\equiv${\boldmath $\mu\cdot H$} is a highest weight representation integrable at level $k$,
taking value in the Cartan subalgebra of the $G$ Lie algebra. This condition is a consequence  of  global issues \cite{Fuchs:2007fw}.
Note that under the full gauge transformation
\be
g_1\mapsto h_1g_1h_1^{-1}\hspace{0.5cm}{\rm and}\hspace{0.5cm} g_2\mapsto h_2g_2h_2
\ee

the parameters in (\ref{qman}) transform as
\bea
&C_{\mu}\mapsto h_1C_{\mu}h_1^{-1}\\ \nonumber
&p\mapsto h_1ph_1^{-1}\\ \nonumber
&L_1\mapsto L_1h_1^{-1}h_2\\ \nonumber
&L_2\mapsto L_2h_1h_2
\eea
Using the Polyakov-Wiegamann identity (\ref{pwwz}) one can check
that the condition (\ref{varka}) is satisfied with the following two-form
\bea\label{varfor}
\varpi(C_{\mu},p, L_1, L_2)= \omega_{\mu}(C_{\mu})-{\rm Tr}(C_{\mu}^{-1}dC_{\mu}dpp^{-1})+{\rm Tr}(p^{-1}dpdL_2L_2^{-1})+\\ \nonumber
+{\rm Tr}(L_1^{-1}dL_1dpp^{-1})+{\rm Tr}(L_1^{-1}dL_1pdL_2L_2^{-1}p^{-1})-
{\rm Tr}(L_1^{-1}dL_1L_2^{-1}dL_2)
\eea
where $\omega_{\mu}(C_{\mu})={\rm Tr}(l^{-1}dle^{2i\pi\mu/k}l^{-1}dle^{-2i\pi\mu/k})$.
Now the full action can be written as
\be\label{facd}
S^{\rm A-V}=S^{\rm kin-def}+S^{\rm gauge-def}+S^{\rm top-def}
\ee
here
\be
S^{\rm kin-def}={k\over 4\pi}\int_{\Sigma_1}dx^+dx^- L^{\rm kin}(g_1)+
{k\over 4\pi}\int_{\Sigma_2}dx^+dx^- L^{\rm kin}(g_2)
\ee
and
\be
S^{\rm gauge-def}={k\over 2\pi }\int_{\Sigma_1}L_v^{\rm gauge}(g_1,A_1)dx^+dx^-
+{k\over 2\pi }\int_{\Sigma_2}L_a^{\rm gauge}(g_2,A_2)dx^+dx^-
\ee
It is cumbersome but possible to check that the action (\ref{facd})
is invariant the gauge transformations:
\bea
g_1\mapsto h_1g_1h_1^{-1}\, ,\hspace{1cm} A_1\mapsto A_1+dh_1h_1^{-1}\\ \nonumber
g_2\mapsto h_2g_2h_2\, ,\hspace{1cm} A_2\mapsto A_2+dh_2h_2^{-1}
\eea
where $h_1: \Sigma_1\rightarrow U(1)$ and $h_2: \Sigma_2\rightarrow U(1)$.

\subsection{Duality defect for the parafermion disc $SU(2)/U(1)$}

Specialize now to the case of $G=SU(2)$  \cite{Bardacki:1990wj}.\\
We write the group elements using the Euler coordinates:
\be
g=e^{i\chi{\sigma_3\over 2}}e^{i\theta\sigma_1}
e^{i\varphi{\sigma_3\over 2}}=e^{i(\tilde{\phi}+\phi){\sigma_3\over 2}}e^{i\theta\sigma_1}
e^{i(\tilde{\phi}-\phi){\sigma_3\over 2}}
\ee
The ranges of the variables are $0\leq\theta\leq{\pi\over 2}$, \hspace{0.2cm}   $0\leq\varphi\leq 2\pi$, \hspace{0.2cm} $0\leq\chi\leq 4\pi$,\hspace{0.2cm}
$-\pi\leq\phi,\tilde{\phi}\leq \pi$.

The axially gauged model ${{\rm SU}(2)\over U(1)}_{\rm axial}$
 is derived by the gauging of the $U(1)$  symmetry
corresponding to shifting of $\tilde{\phi}$ and has the target space
$M_A$ with the following metric and dilaton field \cite{Giveon:1994fu,Maldacena:2001ky}:

\bea\label{axon}
&&ds^2=k(d\theta^2+\tan^2\theta d\phi^2)\\ \nonumber
&&e^{\Phi}={g_s\over \cos\theta}\\ \nonumber
&&\phi\sim \phi+2\pi
\eea
 Using the T-duality rules of the previous  section one can see that T-dual background to the axially gauged model is
\bea\label{tbb}
&&\tilde{ds}^2=k\left(d\tilde{\theta}^2+{d\tilde{\phi}^2\over \tan^2\tilde{\theta}}\right)\\ \nonumber
&&e^{\tilde{\Phi}}={g_s\over \sqrt{k}\sin\tilde{\theta}}\\ \nonumber
&&\tilde{\phi}\sim \tilde{\phi}+{2\pi\over k}
\eea

 Vectorially gauged model ${{\rm SU}(2)\over U(1)}_{\rm vec}$
is derived by the gauging of the $U(1)$ symmetry corresponding to the shifting of $\phi$ and
has the target space $M_V$
with the metric and the dilaton:
\bea\label{vgm}
&&\tilde{ds}^2=k\left(d\tilde{\theta}^2+{d\tilde{\phi}^2\over \tan^2\tilde{\theta}}\right)\\ \nonumber
&&e^{\tilde{\Phi}}={g_s\over \sin\tilde{\theta}}\\ \nonumber
&&\tilde{\phi}\sim \tilde{\phi}+2\pi
\eea
Comparing (\ref{tbb}) and (\ref{vgm}) one can see  that the background T-dual to the axially
gauged model is the $Z_k$ orbifold of the vectorially gauged model.

According to the results of the previous section the world-volume  of the T-duality defect $D^T_{A}$
between backgrounds (\ref{axon}) and (\ref{tbb}) is the submanifold
$\theta=\tilde{\theta}$ of the product $M_V\times M_A$ with the flux $F=d\phi\wedge d\tilde{\phi}$.
The defects between backgrounds (\ref{axon}) and ({\ref{vgm}) $D_{V-A}$ has the same world volume
but the flux is $F=kd\phi\wedge d\tilde{\phi}$.

Consider the defects given by  equation (\ref{qman}).
The conjugacy class takes the form ${C_j=le^{2\pi ij\sigma_3\over k}l^{-1}}$, $j=0,{1\over 2}\ldots {k\over 2}$,
 (since we are working in the specific case of $G=SU(2)$, the general subscript $\mu$ was changed to $j$, which is standard for this group)
and therefore we have a family of the defects labelled by $j$.
Now we show that the T-duality defect above,  $D_{V-A}$, corresponds to $j=0$.

Let us examine this defect in more detail.
Parameterizing $L_1=e^{i\alpha_1\sigma_3/2}$ and $L_2=e^{i\alpha_2\sigma_3/2}$
and writing $p$ using the Euler coordinates, we obtain for this special defect:
\be\label{defsu2}
(g_1, g_2)=\left(e^{i(\tilde{\kappa}+\kappa){\sigma_3\over 2}}e^{i\theta\sigma_1}
e^{i(\tilde{\kappa}-\kappa){\sigma_3\over 2}},\hspace{0.5cm}
e^{i(\tilde{\kappa}+\kappa+\alpha_1){\sigma_3\over 2}}e^{i\theta\sigma_1}
e^{i(\tilde{\kappa}-\kappa+\alpha_2){\sigma_3\over 2}}\right)
\ee
From (\ref{defsu2}) it can be seen that this defect satisfies the condition $\theta=\tilde{\theta}$.
To project down this defect to the product space $M_V\times M_A$ we impose gauge fixing conditions
$\kappa=0$ for the first vectorially gauged model and
\be\label{phta}
(\tilde{\kappa}+\kappa+\alpha_1)+(\tilde{\kappa}-\kappa+\alpha_2)=0
\ee
for the axially gauged model.
From (\ref{phta}) one obtains:
\be\label{kapon}
\tilde{\kappa}=-{\alpha_1+\alpha_2\over 2}
\ee

Therefore the angles $\phi$ and $\tilde{\phi}$ of the target spaces are related to the defect parameters by equations:
\be\label{pha1}
\tilde{\phi}=\tilde{\kappa}=-{\alpha_1+\alpha_2\over 2}
\ee
\be\label{pha2}
\phi={\alpha_1-\alpha_2\over 2}
\ee

Let us evaluate the two-form (\ref{varfor}). For $j=0$ it simplifies to:
\bea\label{varfor2}
\varpi(p, L_1, L_2)= {\rm Tr}(p^{-1}dpdL_2L_2^{-1})
+{\rm Tr}(L_1^{-1}dL_1dpp^{-1})+\\ \nonumber
{\rm Tr}(L_1^{-1}dL_1pdL_2L_2^{-1}p^{-1})-
{\rm Tr}(L_1^{-1}dL_1L_2^{-1}dL_2)
\eea
This implies
\bea
{\rm Tr}(p^{-1}dpdL_2L_2^{-1})=-(d\tilde{\kappa}\cos^2\theta-d\kappa\sin^2\theta)d\alpha_2\\ \nonumber
{\rm Tr}(L_1^{-1}dL_1dpp^{-1})=-d\alpha_1(d\tilde{\kappa}\cos^2\theta+d\kappa\sin^2\theta)\\ \nonumber
{\rm Tr}(L_1^{-1}dL_1pdL_2L_2^{-1}p^{-1})=-d\alpha_1d\alpha_2(\cos^2\theta-{1\over 2})\\ \nonumber
-{\rm Tr}(L_1^{-1}dL_1L_2^{-1}dL_2)={d\alpha_1d\alpha_2\over 2}
\eea

Using that $\kappa=0$ and (\ref{kapon}), (\ref{pha1}) and (\ref{pha2})  one obtains that the $\theta$ dependent terms drop
and we end up with
\be
{k\over 4\pi}\varpi(p, L_1, L_2)={k\over 4\pi}d\alpha_1d\alpha_2={k\over 2\pi}d\tilde{\phi}d\phi
\ee

This is the flux on the defect $D_{V-A}$ and as demonstrated in sec. 2, this defect is topological.

It is shown in  appendix B  that a generic defect  has a geometry given by the inequality
\be\label{insp}
\cos2(\theta-\tilde{\theta})\geq \cos{4\pi j\over k}
\ee

\subsection{Axial-vectorial defects as operators in the parafermion Hilbert space}

It  has been shown that the backgrounds (\ref{axon}) and (\ref{vgm}) correspond to the parafermion theory,
and therefore the defects above  can be realized as operators in the parafermions Hilbert space.

To construct the corresponding operator one should start
with the Cardy defect in the parafermion theory \cite{Petkova:2000ip}:
\be\label{defc}
X_{\hat{j},\hat{n}}=\sum_{j,n}{S^{\rm PF}_{(\hat{j},\hat{n});(j,n)}\over S^{\rm PF}_{(0,0);(j,n)}}
P^{\rm PF}_{j,n}\bar{P}^{\rm PF}_{j,n}
\ee
Here $S^{\rm PF}_{(\hat{j},\hat{n});(j,n)}$ is the parafermion matrix of the modular transformation
\be
S^{\rm PF}_{(\hat{j},\hat{n});(j,n)}=\sqrt{2\over k}S^{{\rm SU}(2)}_{\hat{j}j}e^{i\pi n\hat{n}\over k}
\ee
$P^{\rm PF}_{j,n}$ and $\bar{P}^{\rm PF}_{j,n}$ are projectors
\be
P^{\rm PF}_{j,n}=\sum_N|j,n, N\rangle_0\otimes_1\langle j,n,N|
\ee
\be
\bar{P}^{\rm PF}_{j,n}=\sum_M\overline{|j,n, M\rangle_0}\otimes_1\overline{\langle j,n,M|}
\ee
where the sums over $M$ and $N$ are over orthonormal bases of the parafermion state spaces.
Subscriptes $0$ and $1$ distinguish between the theories on the two sides of the defect.
Here $j\in\{0,{1\over 2},\ldots {k\over 2}\}$ and $n\in \mathbb{Z}/2k\mathbb{Z}$ satisfy the constraint
$2j+n=0\; {\rm mod} \; 2$. The pairs $(j,n)$ and $(k/2-j, k+n)$  have to be identified.
We need to construct a defect mapping $A$- branes to $B$- branes.
This can be done along the lines used in \cite{Maldacena:2001ky} for the parafermion $B$-
branes construction.
Recall that the $\mathbb{Z}_k$ orbifold of the parafermion theory at level $k$
is T-dual to the original theory.
To get a defect mapping $A$- branes to $B$- branes one should sum over $\mathbb{Z}_k$
images of $X_{\hat{j},\hat{n}}$ and perform T-duality. In order to circumvent the fixed point problem, we consider the case of odd $k$
\footnote{In the case of an even $k$, the primary field ${k\over 4}$ has the non-trivial stabilizator $\mathbb{Z}_2$, which requires the fixed point resolution procedure.
As a consequence the formulae for branes and defects derived in this way get modified. See for details \cite{Maldacena:2001ky,Brunner:2000nk,Fuchs:2000fd}.}
 .
Summing over images
leaves in (\ref{defc}) only the $n=0$ term and T-duality exchanges $\bar{P}^{\rm PF}_{j,n}$
with its B-type version, which can be derived in the following way.
Define also corresponding projectors for $SU(2)$ :
\be
P^{{\rm SU}(2)}_{j}=\sum_N|j,N\rangle_0\otimes_1\langle j,N|
\ee
\be
\bar{P}^{{\rm SU}(2)}_{j}=\sum_M\overline{|j, M\rangle_0}\otimes_1\overline{\langle j,M|}
\ee
where the sums over $N$ and $M$ are over orthonormal bases of the ${\rm SU}(2)$ state spaces, and rational $U(1)$ scalar:
\be
P^{{\rm U}(1)}_{r\pm}=\exp\left[\pm\sum_{n=1}^{\infty}{\alpha^0_{-n}\alpha^1_n\over n}\right]
\sum_{l\in Z}|{r+2kl\over \sqrt{2k}}\rangle_0\otimes _1\langle\pm{r+2kl\over \sqrt{2k}}|
\ee
\be
\bar{P}^{{\rm U}(1)}_{r'\pm}=\exp\left[\pm\sum_{n=1}^{\infty}{\bar{\alpha}^0_{-n}\bar{\alpha}^1_n\over n}\right]
\sum_{l\in Z}\overline{|\pm{r'+2kl'\over \sqrt{2k}}\rangle}_0\otimes _1\overline{\langle{r'+2kl'\over \sqrt{2k}}|}
\ee
Using the decomposition of $SU(2)_k$ as a product of parafermion and scalar theories one can write
\be
\bar{P}^{{\rm SU}(2)}_{j}=\sum_{r}\bar{P}^{\rm PF}_{j,r}\bar{P}^{{\rm U}(1)}_{r+}
\ee
To define the T-dual projector $\overline{BP}^{\rm PF}_{j,n}$ we rotate the $SU(2)$ projector
$\bar{P}^{{\rm SU}(2)}_{j}$ with operator $e^{i\pi \bar{J}^1_0}$, satisfying
\be
e^{i\pi \bar{J}^1_0} \bar{J}^3_0e^{-i\pi \bar{J}^1_0}=-\bar{J}^3_0
\ee
and afterwards decompose it again as a product of the parafermion and scalar theories:
\be
1\otimes e^{i\pi \bar{J}^1_0}\bar{P}^{{\rm SU}(2)}_{j}=\sum_{r}\overline{BP}^{\rm PF}_{j,r}\bar{P}^{{\rm U}(1)}_{r-}
\ee
Combining the orbifolding and the T duality  procedures results is:
\be\label{opdef}
Y^{AB}_{\hat{j}}=\sqrt{k}\sum_{j}{S^{{\rm SU}(2)}_{\hat{j},j}\over S^{{\rm SU}(2)}_{0,j}}
P^{\rm PF}_{j,0}\overline{BP}^{\rm PF}_{j,0}
\ee
It is shown in the appendix that in the large $k$ limit $Y^{AB}_{j}$ has the geometry given with the overlap

\bea\label{ovlapon26}
&& \langle\theta, \phi|Y^{AB}_{\hat{j}}|\tilde{\theta},\tilde{\phi}\rangle\sim \\ \nonumber
 && {k\over \pi^2}\int_{|2\theta-2\tilde{\theta}|}^{2\theta+2\tilde{\theta}}
{\Theta(\cos\gamma-\cos2\hat{\psi})\over
\sqrt{\cos\gamma-\cos2\hat{\psi}}}
{\sin \gamma d\gamma\over \sqrt{[\cos\gamma-\cos2(\theta+\tilde{\theta})][\cos2(\theta-\tilde{\theta})-\cos\gamma]}}
\eea
where $\hat{\psi}={(2\hat{j}+1)\pi\over k+2}$ and $\Theta$ is the Heavyside step function.
Eq. (\ref{ovlapon26}) shows that the world-volume of the defect
should satisfy the inequality
\be
\cos2(\theta-\tilde{\theta})\geq \cos\hat{\psi}
\ee

which in the large $k$ limit coincides with the inequality (\ref{insp}), defining the geometry of a generic defect.

Note that in the defect $Y^{AB}_0$, the relation of the elements of the matrix of the modular transformation drops, and it  is a sum of projectors,
projecting down to the $n=0$ subspace and performing T-duality, thus
mapping the $A_{j,n}$ Cardy branes to the $B_j$ branes constructed in \cite{Maldacena:2001ky}. For generic $\hat{j}$  one derives a linear
combination of  the $B_l$ branes with coefficients given by the fusion numbers $N_{\hat{j}j}^l$.


\section{Fermionic T-duality}
In this section we show how do defects generate T-duality on fermionic coordinates.
We show here that the fermionic T-duality is implemented by the defect, given by the fermionic analogue of the Poincar\'{e} line bundle, which we
call Super-Poincar\'{e} line bundle. This defect is invertible.

Then we define the  super Fourier-Mukai  transform, as in the bosonic case, as an integral with an appropriate kernel given by the exponent of the  flux of a super Poincare line bundle. \\

\subsection{Pseudodifferential forms integration}

The technical details can be found in appendix C,  the result \cite{berlei1,berlei2,lavaud} is presented here. Pseudodifferential forms, defined on a supermanifold of $p$ bosonic and $q$ fermionic coordinates,  are of the form
\be f=\sum_{v,u}{f_{v,u}(x,d\theta)\theta^v dx^u} \label{pseudodiff1} \ee
Where: $v={v_1,...,v_q}$; $u={u_1,...,u_p}$; $v_i,u_i\in {0,1}$; $x=x_1, ... x_p$; $d\theta = d\theta_1, ... d\theta_q$;
$\theta^v = \theta_1^{v_1}\cdot ... \cdot \theta_q^{v_q}$; $dx^u = dx_1^{u_1}\cdot ... \cdot dx_p^{u_p}$, and the sum is over all possible values of $u$ and $v$. Such an object can be integrated over the bundle on which it is defined. The integration  is defined as
\be \int_\mathcal{{B}}f = \int_B f_{1,1,...,1}\label{superint1} \ee
Where $\mathcal{B}$ is the cotangent bundle of the supermanifold and $B$ is its underlying bundle, with just the bosonic coordinates. The $d\theta$s are coordinates along the bundle, and unlike the case of the fibrewise integration presented above, they are bosonic. For that reason one needs $f$ to be sufficiently rapidly decreasing  in them in order for the integral to converge. As will be demonstrated bellow, this is indeed the case for the super Fourier-Mukai transform.\\

\subsection{Review of the fermionic T-duality}

Consider  the action (\ref{actbg}) for the case when one has fermionic as well as bosonic variables, and $G_{ij}$ and $B_{ij}$
are graded-symmetric and graded -antisymmetric tensors respectively. Suppose that $G_{ij}$ and $B_{ij}$ do not depend on the fermionic
variable $\theta^1$ \cite{Berkovits:2008ic}. Separating the variable  $\theta^1$ one has

\be\label{actfer}
S=\int dx^+dx^- (B_{11}\partial \theta^1\bar{\partial}\theta^1+E_{1N}\partial \theta^1\bar{\partial}X^N+E_{M1}\partial X^M\bar{\partial}\theta^1+E_{MN}\partial X^M\bar{\partial}X^N)
\ee
 Replacing derivatives of  $\theta^1$ by fermionic vector $(A,\bar{A})$ and introducing a Lagrange multiplier field $\tilde{\theta}^1$ one gets
\be\label{tutu}
S=\int dx^+dx^-(B_{11}A\bar{A}+E_{1N}A\bar{\partial}X^N+E_{M1}\partial X^M\bar{A}+E_{MN}\partial X^M\bar{\partial}X^N+\tilde{\theta}^1(\partial \bar{A}-\bar{\partial}A))
\ee
Integrating out $\tilde{\theta}^1$ imposes that
\be\label{ff1}
A=\partial\theta^1\hspace{0.5cm} {\rm and} \hspace{0.5cm} \bar{A}=\bar{\partial}\theta^1.
\ee
Integrating out $(A,\bar{A})$ results in:
\be\label{ff2}
\bar{A}={1\over B_{11}}\left((-)^{s_M}E_{1M}\bar{\partial}X^M+\bar{\partial}\tilde{\theta}^1\right)\hspace{0.5cm} {\rm and} \hspace{0.5cm}
A=-{1\over B_{11}}\left(E_{M1}\partial X^M-\partial\tilde{\theta}^1\right)
\ee
Inserting (\ref{ff2}) in (\ref{tutu}) one obtains fermionic T-dual background:
\bea\label{ftdual}
&&\tilde{B}_{11}=-{1\over B_{11}}\\ \nonumber
&&\tilde{E}_{1M}={E_{1M}\over B_{11}}\\ \nonumber
&&\tilde{E}_{M1}={E_{M1}\over B_{11}}\\ \nonumber
&&\tilde{E}_{MN}=E_{MN}-{E_{1N}E_{M1 }\over B_{11}}
\eea
or in the components:
\bea\label{ftdualc}
&&\tilde{B}_{11}=-{1\over B_{11}}\\ \nonumber
&&\tilde{G}_{1M}={G_{1M}\over B_{11}}\\ \nonumber
&&\tilde{B}_{1M}={B_{1M}\over B_{11}}\\ \nonumber
&&\tilde{G}_{MN}=G_{MN}-{1\over B_{11}} (G_{1N}B_{M1}+B_{1N}G_{M1})\\ \nonumber
&&\tilde{B}_{MN}=B_{MN}-{1\over B_{11}} (G_{1N}G_{M1}+B_{1N}B_{M1})
\eea
Equating (\ref{ff1}) and (\ref{ff2}) one gets:
\be\label{ftsca}
\partial\tilde{\theta}^1=B_{11}\partial\theta^1+E_{M1}\partial X^M\hspace{0.5cm} {\rm and} \hspace{0.5cm}
\bar{\partial}\tilde{\theta}^1=B_{11}\bar{\partial}\theta^1-(-)^{s_M}E_{1M}\bar{\partial} X^M
\ee
The rest of the coordinates remains  unchanged.

\subsection{Defects implementing the fermionic T-duality and the Super Poincar\'{e} line bundle}
We now consider the action with defect, with target spaces related by the equations (\ref{ftdual}), and the defect given again
by the correspondence space
\be
X^N=\tilde{X}^N,\hspace{1cm} N=2\ldots {\rm dim} M
\ee
 and connection
\be
A=\theta^1d\tilde{\theta}^1
\ee
 with curvature
\be
F=d\theta^1\wedge d\tilde{\theta}^1.\ee
We will call this super line bundle by analogy with the bosonic case a Super-Poincar\'{e}  bundle.
Now the defect  equations of motion take the form:
\be\label{ftd1}
E_{j1}\partial X^j-(-)^{s_j}E_{1j}\bar{\partial} X^j-\partial_{\tau}\tilde{\theta}^1=0
\ee

\be\label{ftd2}
E_{jN}\partial X^j-(-)^{s_js_N}E_{Nj}\bar{\partial} X^j-\tilde{E}_{jN}\partial \tilde{X}^j+(-)^{s_js_N}\tilde{E}_{Nj}\bar{\partial} \tilde{X}^j=0,\hspace{0.5cm} N=2\ldots {\rm dim} M
\ee

\be\label{ftd3}
\tilde{E}_{j1}\partial \tilde{X}^j-(-)^{s_j}\tilde{E}_{1j}\bar{\partial} \tilde{X}^j+\partial_{\tau}\theta^1=0
\ee

Additionally as before we have:
\be\label{fdxt}
\partial X^N+\bar{\partial} X^N=\partial \tilde{X}^N+\bar{\partial} \tilde{X}^N,\hspace{0.5cm} N=2\ldots {\rm dim} M
\ee
Solving (\ref{ftd1}),  (\ref{ftd2}), (\ref{ftd3}), (\ref{fdxt}) we obtain

\bea\label{defsolf}
&&\bar{\partial} \tilde{X}^N=\bar{\partial} X^N,\hspace{1cm} N=2\ldots {\rm dim} M\\ \nonumber
&&\partial \tilde{X}^N=\partial X^N,\hspace{1cm} N=2\ldots {\rm dim} M\\ \nonumber
&&\partial\tilde{\theta}^1=B_{11}\partial\theta^1+E_{M1}\partial X^M\\ \nonumber
&&\bar{\partial}\tilde{\theta}^1=B_{11}\bar{\partial}\theta^1-(-)^{s_M}E_{1M}\bar{\partial} X^M
\eea
The details of the calculation can be found in appendix D.
The relations (\ref{defsolf}) coincide with the equations (\ref{ftsca}). Therefore the defect given
by the Super-Poincare bundle on the super-correspondence space induces the fermionic T-duality.

One can check that equations (\ref{ftdualc}) and (\ref{defsolf}) imply:
\be
T=G_{ij}\partial  X^i\partial  X^j=\tilde{T}=\tilde{G}_{ij}\partial \tilde{X}^i\partial \tilde{X}^j
\ee
and
\be
\bar{T}=G_{ij}\bar{\partial} X^i\bar{\partial} X^j=\tilde{\bar{T}}=\tilde{G}_{ij}\bar{\partial} \tilde{X}^i\bar{\partial} \tilde{X}^j
\ee
which means that the defect is topological.

All this again can be  generalized to the T-dualizing of  several coordinates.
Suppose we T-dualize the first $n$ coordinates, indexed by Greek letters.

The transformed background has the form
\be\label{trgenn}
\tilde{E}=\left(\begin{array}{ccc}
-E^{-1}_{\alpha\beta}& E^{-1}_{\alpha\beta}E_{\beta N}\\
E_{M\alpha}E^{-1}_{\alpha\beta}&E_{MN}-E_{\beta N}E_{M\alpha} E^{-1}_{\alpha\beta}\end{array}\right)
\ee

Now we should consider the defect with the worldvolume
\be
X^N=\tilde{X}^N,\hspace{1cm} N=n+1\ldots {\rm dim} M
\ee
 and connection
\be
A=\sum_{\alpha=1}^n\theta^{\alpha}d\tilde{\theta}^{\alpha}.
\ee
 It has the curvature
\be
F=\sum_{\alpha=1}^nd\theta^{\alpha}\wedge d\tilde{\theta}^{\alpha}.
\ee
In the same way as above we can show that for $M$ and $\tilde{M}$ related by equations (\ref{trgenn}) this defect is topological and implies the defect equations of motion:
\bea
&&\bar{\partial} \tilde{X}^N=\bar{\partial} X^N,\hspace{1cm} N=n+1\ldots {\rm dim} M\\ \nonumber
&&\partial \tilde{X}^N=\partial X^N,\hspace{1cm} N=n+1\ldots {\rm dim} M\\ \nonumber
&&\partial \tilde{\theta}^{\alpha}=E_{\beta\alpha}\partial \theta^{\beta}+E_{M\alpha}\partial X^M\\ \nonumber
&&\bar{\partial} \tilde{\theta}^{\alpha}=E_{\alpha\beta}\bar{\partial} \theta^{\beta}-(-)^{s_M}E_{\alpha M}\bar{\partial} X^N
\eea
We have  obtained  again T-duality relations for  several T-dualized fermionic coordinates.
\subsection{Super Fourier-Mukai transform}

We now elaborate the Fourier-Mukai transform for  fermionic  T-duality.
It has the form:
\be\label{fmtf}
e^{-\hat{B}}\hat{\cal G}=\int d\eta e^{-B}{\cal G}e^{\eta \tilde{\eta}}
\ee
with ${\cal G}$ and $B$ as in (\ref{fmt}), where we set $\eta=d\theta^1$.
As we explained $\eta$ is  a bosonic variable, so we have a usual integration over $\eta$.
From (\ref{ftdualc}) one obtains:
\bea
&& \hat{B}-B=-{1\over 2B_{11}}\tilde{\eta}^2-{1\over 2}B_{11}\eta^2 \\ \nonumber
&& -{1\over 2B_{11}} (G_{1N}G_{M1}+B_{1N}B_{M1})dX^MdX^N+
{B_{1M}\over B_{11}}\tilde{\eta} dX^M-B_{1M}\eta dX^M
\eea

Suppose that ${\cal G}$ does not depend on $\eta$.
Using the formula for the Gaussian integral
\be\label{gauss}
\int dx e^{-{1\over 2}ax^2+Jx}={\sqrt{2\pi}\over \sqrt{a}}e^{{J^2\over 2a}}
\ee
we obtain that the terms in (\ref{fmtf}) containing $B_{1M }$ and the first quadratic term are canceled and, we end up with
\be\label{gb11}
\hat{\cal G}={\sqrt{2\pi}\over \sqrt{B_{11}}}{\cal G}e^{-{1\over 2B_{11}} G_{1N}G_{M 1}dX^{M}dX^{N}}
\ee
Note that $G_{1N}$ and $B_{1N}$ have parity $(-)^{s_N+1}$.
Hence if $dX^{M}$ and $dX^{N}$ are differentials of the bosonic coordiantes,
the product $G_{1N}G_{M 1}$ contains fermionic coordinates and drops if we consider the lowest
$\theta=0$ components,
in agreement with the observation \cite{Berkovits:2008ic} that the fermionic T-duality does not modify D-brane dimensionality.
Note that the lowest $\theta=0$ components of (\ref{gb11}) coincide with the homogeneous part  of the transformation of the Ramond-Ramond forms in \cite{Berkovits:2008ic}.

Using the transformations rules (\ref{trgenn}) equation (\ref{gb11})   can be generalized  to the case
of the T-dualization of  several fermionic variables $\theta^{\alpha}$.
Keeping in mind that eventually we are going to project to the $\theta=0$ component
we can set $G_{\alpha\beta}=0$, since
$G_{\alpha\beta}=\eta^{ab}E_a^{\alpha}E_b^{\beta}$, and taking into account
that $a$ and $b$ are bosonic and $\alpha$ and $\beta$ are fermionic, one sees that
$E_a^{\alpha}$ and $E_b ^{\beta}$ are odd.
With this simplification the Fourier-Mukai transform for ${\cal G}$ independent on $\theta^{\alpha}$ can be computed to yield:
\be\label{gb11sv}
\hat{\cal G}={\sqrt{2\pi}\over \sqrt{{\rm det}||B_{\alpha\beta}||}}{\cal G}e^{-{1\over 2} B_{\alpha\beta}^{-1}G_{\alpha N}G_{M \beta}dX^{M}dX^{N}}
\ee
The lowest component of  (\ref{gb11sv})  again coincides with the  homogeneous part
of the transformation of Ramond-Ramond  forms in \cite{Berkovits:2008ic} for the fermionic T-dualization of the $n$ coordinates.

\section{Discussion}
It is shown in \cite{Fre:2009ki} that the generalization of the $SO(d,d)$ duality group for the sigma models with a super target space
is the orthosymplectic group $OSp(d,d|2n)$. As in the bosonic case this group is generated by the superspace field redefinitions, the super B-field shift
and the bosonic and fermionic  dualities. Therefore the corresponding defects are given by the bosonic and fermionic dualities defects, constructed in
sections 2.2 and 4.3 correspondingly, and the superspace analogue of the diagonal defects constructed in section 2.3.
 Some of these defects, as in the bosonic case considered in \cite{Bachas:2012bj}, can be non-invertible.
Their study can lead to a new class of interfaces and is left for future work.
The entries of the (semi)-group of defects  should be found from the analysis of the admissibility of the corresponding fields and angles of the defects.

It is also an interesting problem to find operator realization of the defect given by the Super-Poincar\'{e} line bundle implementing the
fermionic T-duality.

Another open problem is to identify  the possible connection between  topological defects and the so called T folds \cite{Hull:2004in} and generalize their construction.

\section*{Acknowledgments}
The work of S. Elitzur is partially supported by the Israel Science Foundation Center of Excellence.\\
The work of E. Rabinovici is partially supported by the American-Israeli Bi-National Science Foundation and the Israel Science Foundation Center of Excellence.\\
The work of G. Sarkissian was partially supported by the Research project 11-1c258 of
the State Committee of Science of Republic of Armenia and ANSEF hepth-2774 grant.\\
We would like also to thank Christoph Schweigert and Ingo Runkel for discussions.
\vspace*{3pt}


\appendix
\section{Solution of the defect equations of motion in the bosonic case}
In subsection 2.2 we obtained the following defect equations of motion:
\be\label{td1a}
E_{j1}\partial X^j-E_{1j}\bar{\partial} X^j-\partial_{\tau}\tilde{X}^1=0
\ee

\be\label{td2a}
E_{jN}\partial X^j-E_{Nj}\bar{\partial} X^j-\tilde{E}_{jN}\partial \tilde{X}^j+\tilde{E}_{Nj}\bar{\partial} \tilde{X}^j=0
\ee

\be\label{td3a}
\tilde{E}_{j1}\partial \tilde{X}^j-\tilde{E}_{1j}\bar{\partial} \tilde{X}^j-\partial_{\tau}X^1=0.
\ee

\be\label{dxta}
\partial X^N+\bar{\partial} X^N=\partial \tilde{X}^N+\bar{\partial} \tilde{X}^N
\ee
The index $j$ runs from 1 to ${\rm dim} M$. The capital latin indices run from 2 to  ${\rm dim} M$.

To solve these equations we perform the following steps.

Separating the first coordinate in (\ref{td1a}) and (\ref{td3a}) we obtain
\be\label{e1a}
E_{11}(\partial X^1-\bar{\partial} X^1)+E_{M1}\partial X^M-E_{1M}\bar{\partial} X^M-\partial \tilde{X}^1-\bar{\partial} \tilde{X}^1=0
\ee
\be\label{e2a}
-E_{11}(\partial X^1+\bar{\partial} X^1)-E_{M1}\partial \tilde{X}^M-E_{1M}\bar{\partial} \tilde{X}^M+\partial \tilde{X}^1-\bar{\partial} \tilde{X}^1=0
\ee
Taking sum and difference of (\ref{e1a}) and (\ref{e2a}) and taking into account (\ref{dxta}) one gets
\be\label{ex1a}
E_{11}\partial X^1-\partial \tilde{X}^1+E_{M1}\partial \tilde{X}^M+G_{M1}(\bar{\partial} \tilde{X}^M-\bar{\partial} X^M)=0
\ee
\be\label{ey1a}
-E_{11}\bar{\partial} X^1-\bar{\partial} \tilde{X}^1-E_{1M}\bar{\partial} \tilde{X}^M+G_{M1}(\bar{\partial} \tilde{X}^M-\bar{\partial} X^M)=0
\ee
Separating the first coordinate in (\ref{td2a}) and again  using (\ref{dxta}) we receive
\bea\label{exya}
&& {E_{1N}\over E_{11}}\left (E_{11}\partial X^1-\partial \tilde{X}^1+E_{M1}\partial \tilde{X}^M\right) \\ \nonumber 
&& -{E_{N1}\over E_{11}}\left(E_{11}\bar{\partial} X^1+\bar{\partial} \tilde{X}^1+E_{1M}\bar{\partial} \tilde{X}^M\right)+ 2G_{MN}\left(\bar{\partial} \tilde{X}^M-\bar{\partial} X^M\right)=0
\eea

Combining (\ref{ex1a}), (\ref{ey1a}) and (\ref{exya}) finally we reach  the equations
\bea
&&\bar{\partial} \tilde{X}^N=\bar{\partial} X^N\hspace{1cm} N=2,\ldots {\rm dim}M\\ \nonumber
&&\partial \tilde{X}^N=\partial X^N\hspace{1cm} N=2,\ldots {\rm dim}M\\ \nonumber
&&\partial \tilde{X}^1=E_{11}\partial X^1+E_{M1}\partial X^M\\ \nonumber
&&\bar{\partial} \tilde{X}^1=-(E_{11}\bar{\partial} X^1+E_{1M}\bar{\partial} X^M)
\eea

\section{Geometry of the vector-axial  duality defects for $SU(2)/U(1)$ cosets}
Here we provide details of the computation of the geometry of the defects considered in section 3.
It is interesting to note that geometrically ( but not the flux and the symmetries! ) they coincide  with some folded brane considered in \cite{Sarkissian:2003yw}, and one can use the results there.
 For reader convenience we collected the necessary stuff in this appendix.
\be\label{defsu22}
(g_1, g_2)=\left(C_{\mu}p
,\hspace{0.2cm}
e^{i\alpha_1{\sigma_3\over 2}}p
e^{i\alpha_2{\sigma_3\over 2}}\right)
\ee
The conjugacy class is ${C_j=le^{2\pi ij\sigma_3\over k}l^{-1}}$.

Equation (\ref{defsu22})  implies

\be\label{g1g2}
{\rm Tr}(g_1e^{i\alpha_2{\sigma_2\over 2}}g_2^{-1}e^{i\alpha_1{\sigma_2\over 2}})=2\cos{2j\pi\over k}
\ee
Consider (\ref{g1g2}) as an equation in $\alpha_1$ and $\alpha_2$.
The question is,  which condition $g_1$ and $g_2$ should satisfy, in order that
(\ref{g1g2}) has solutions in $\alpha_1$ and $\alpha_2$.
To answer this question we introduce a new element
$F=g_1e^{i\alpha_2{\sigma_2\over 2}}g_2^{-1}$ and
analyze first for which $F$ there exists an $\alpha_1$ solving the equation
\be
{\rm Tr}(Fe^{i\alpha_1{\sigma_2\over 2}})=2\cos{2j\pi\over k}
\ee
Denoting the Euler coordinates of $F$ by $\theta_F$, $\tilde{\phi}_F$ and $\phi_F$,
this equation takes the form
\be\label{tfc}
\cos\theta_F\cos(\tilde{\phi}_F+\alpha_1/2)=\cos{2j\pi\over k}
\ee
Eq. (\ref{tfc}) has solution in $\alpha_1$ only if the inequality
\be
\cos2\theta_F\geq \cos{4j\pi\over k}
\ee
is satisfied.
Using the formula for the Euler angles of the product of two elements \cite{vilenkin}
\be
\cos2\hat{\theta}=\cos2\theta_1\cos2\theta_2-\sin2\theta_1\sin2\theta_2\cos(\chi_2+\varphi_1)
\ee

we obtain:
\be\label{coss}
\cos2\theta_1\cos2\theta_2+\sin2\theta_1\sin2\theta_2\cos(\varphi_1-\varphi_2+\alpha_2)
\geq \cos{2j\pi\over k}
\ee
For the inequality (\ref{coss}) to have solution in $\alpha_2$, the maximum value of the left hand side
should be greater than  $\cos{2j\pi\over k}$. The maximum value of the left hand side
is $\cos2(\theta_1-\theta_2)$.
Therefore for (\ref{coss}) to have solutions the inequality
\be
\cos2(\theta_1-\theta_2)\geq \cos{4j\pi\over k}
\ee
should be satisfied.

Now we turn to the calculation of the geometry of the defect corresponding to the operator
(\ref{opdef}).

The matrix of the modular transformation of the $SU(2)$ WZW model at the level $k$ is
\be
S^{{\rm SU}(2)}_{\hat{j}j}=\sqrt{2\over k+2}\sin\left({(2\hat{j}+1)(2j+1)\over k+2}\right)
\ee
In the large -$k$ limit the ratio of the $S$ matrix elements appearing in the defect operator
simplifies to
\be
{S^{{\rm SU}(2)}_{\hat{j}j}\over S^{{\rm SU}(2)}_{0j}}\sim
{k\over \pi(2j+1)}\sin[(2j+1)\hat{\psi}]
\ee
where we have introduced $\hat{\psi}={(2\hat{j}+1)\pi\over k+2}$.
To compute the overlap of the defect with the eigen-position state, we should remember that the coordinate wave functions
of the parafermion theory are  given by a gauge invariant  wave function on $SU(2)$. Gauge invariance  means here that the wave functions
 are independent of the Euler angle $\phi$ or $\tilde{\phi}$   in the axially  or vectorilally gauged models correspondingly.
On the other  hand wave functions on $SU(2)$ are given by the normalized Wigner functions $\sqrt{2j+1}{\cal D}^j_{nm}$ \cite{Maldacena:2001ky}.
Therefore in the axially gauged model the  wave functions are $\sqrt{2j+1}{\cal D}^j_{m,-m}$ and in the vectorially gauged model they are
$\sqrt{2j+1}{\cal D}^j_{mm}$. In the defect (\ref{opdef})  only  modes with $m=0$ are present. Remembering that ${\cal D}^j_{00}$
are the Legendre polynomials, finally we obtain
at the large $k$ level :
\be
\langle\theta, \phi|Y^{AB}_{\hat{j}}|\tilde{\theta},\tilde{\phi}\rangle=
\sum_j{k\over \pi}\sin[(2j+1)\hat{\psi}]P_j(\cos2\theta)P_j(\cos2\tilde{\theta})
\ee
where $P_j$ are the Legendre polynomials.
Using the formula \cite{vilenkin}:
\bea
&& P_j(\cos\theta_1)P_j(\cos\theta_2)= \\ \nonumber 
&& {1\over \pi}\int_{|\theta_1-\theta_2|}^{\theta_1+\theta_2}
P_j(\cos\gamma){\sin \gamma d\gamma\over \sqrt{[\cos\gamma-\cos(\theta_1+\theta_2)][\cos(\theta_1-\theta_2)-\cos\gamma]}}
\eea
we obtain:
\bea\label{ovlapon}
&& \langle\theta, \phi|Y^{AB}_{\hat{j}}|\tilde{\theta},\tilde{\phi}\rangle= \\ \nonumber
&& {k\over \pi^2}\int_{|2\theta-2\tilde{\theta}|}^{2\theta+2\tilde{\theta}}\sum_j\sin[(2j+1)\hat{\psi}]
P_j(\cos\gamma){\sin \gamma d\gamma\over \sqrt{[\cos\gamma-\cos2(\theta+\tilde{\theta})][\cos2(\theta-\tilde{\theta})-\cos\gamma]}}
\eea
Now let us evaluate \cite{Maldacena:2001ky}
\be\label{lgp}
\sum_j\sin[(2j+1)\hat{\psi}]P_j(\cos\gamma)={e^{i\hat{\psi}}\over 2i}\sum_j
e^{(2j)i\hat{\psi}}P_j(\cos\gamma)-{e^{-i\hat{\psi}}\over 2i}\sum_j
e^{-(2j)i\hat{\psi}}P_j(\cos\gamma)
\ee
Using the generating function for Legendre polynomials:
\be
\sum_nt^n P_n(x)={1\over \sqrt{1-2tx+t^2}}
\ee
one obtains for (\ref{lgp}):
\be
{e^{i\hat{\psi}}\over 2i}{1\over \sqrt{e^{2i\hat{\psi}}\left(e^{-2i\hat{\psi}}-2\cos\gamma+e^{2i\hat{\psi}}\right)}}+{\rm c.c.}
=-{1\over 2\sqrt{2}}{1\over \sqrt{\cos\gamma-\cos2\hat{\psi}}}+{\rm c.c.}
\ee
and introducing the Heavyside step function $\Theta$ one gets:
\be\label{heavy}
\sum_j\sin[(2j+1)\hat{\psi}]P_j(\cos\gamma)\sim {\Theta(\cos\gamma-\cos2\hat{\psi})\over
\sqrt{\cos\gamma-\cos2\hat{\psi}}}
\ee
Inserting (\ref{heavy}) in(\ref{ovlapon})  one derives
\bea\label{ovlapon2}
&& \langle\theta, \phi|Y^{AB}_{\hat{j}}|\tilde{\theta},\tilde{\phi}\rangle\sim \\ \nonumber
&& {k\over \pi^2}\int_{|2\theta-2\tilde{\theta}|}^{2\theta+2\tilde{\theta}}
{\Theta(\cos\gamma-\cos2\hat{\psi})\over
\sqrt{\cos\gamma-\cos2\hat{\psi}}}
{\sin \gamma d\gamma\over \sqrt{[\cos\gamma-\cos2(\theta+\tilde{\theta})][\cos2(\theta-\tilde{\theta})-\cos\gamma]}}
\eea
Eq. (\ref{ovlapon2}) shows that the world-volume of the defect
should satisfy the inequality
\be
\cos2(\theta-\tilde{\theta})\geq \cos2\hat{\psi}
\ee
For $\hat{j}=0$ in the large $k$ limit it yields $\theta=\tilde{\theta}$.

\section{Some Super Geometry}
Here we review some of the definitions and results of super geometry, necessary for our needs, and in particular fibre-wise integration on super fibre bundles.
We denote by $\Lambda(m)$ the exterior algebra in $m$ variables over a field $F$. This algebra is generated by an orthonormal basis ${\theta_1,...,\theta_m}$ of $F^m$ and the relations $\theta_i\theta_j=-\theta_j\theta_i$.\\
Next we need to define the notion of a sheaf of objects in a category $\mathcal{C}$ on a space $X$.
First, a presheaf, $\mathcal{F}$, of objects in a category $\mathcal{C}$ on a topological space $X$ is defined such that for every open set $U\subseteq X$ there is an object $\mathcal{F}\in obj(\mathcal{C})$ and for every $V\subseteq U$ there is a morphism $r_{U,V}\in Mor_{\mathcal{C}(\mathcal{F}(U),\mathcal{F}(U))}$, called restriction, with the following conditions:
\begin{enumerate}
\item
$\mathcal{F}(\o)=0$ ($\o$ being the null set, and 0 is a trivial object in $obj(\mathcal{C})$).
\item
$r_{U,U}$ is the identity map between $\mathcal{F}(U)$ and itself.
\item
For $W\subseteq V \subseteq U$, $r_{W,U}=r_{W,V}\circ r_{V,U}$
\end{enumerate}
Now a sheaf is a presheaf with the following added conditions:
\begin{enumerate}
\item
For any open set $U$ and a covering of it ${U_i}$, if $s\in \mathcal{F}(U)$ such that $r_{U,U_i}(s)=0$ for all $i$, then $s=0$.
\item
For any open set $U$ and a covering of it ${U_i}$, if there exists elements $s_i\in\mathcal{F}(U_i)$ for any $i$, such that for any $i,j$,
 $r_{U_i,U_i\cap U_j}(s_i)=r_{U_j,U_i\cap U_j}(s_j)$, then there exists $s\in\mathcal{F}(U)$ such that $r_{U,U_i}=s_i$
\end{enumerate}
One can verify that the algebra of $C^\infty$ function on a metric space $X$ is a sheaf on it.\\
For a metric space $X$, the sheaf of $C^\infty$ functions on $X$ is denoted by $\mathcal{O}_X$. The ring of functions restricted to a subset $U\subset X$ is denoted by $\mathcal{O}_X(U)$.\\
A smoothed superspace, $\mathcal{K}^{p,q}=(k^p,\mathcal{O}_{k^{p,q}})$ is defined as a vector space $k^p$ with a sheaf $\mathcal{O}_{k^{p,q}}$ that is defined by $\mathcal{O}_{k^{p,q}}(U)=\mathcal{O}_{k^p}(U)\otimes\Lambda(q)$.\\
For a domain $U\subset k^p$ one defines a superdomain $\mathcal{U}^{p,q}=(U,\mathcal{O}_{k^{p,q}}|_U)$. $U$ is called the underlying domain of $\mathcal{U}^{p,q}$. If $(x_1,...,x_p)$ are the coordinates of $U$ and $(\theta_1,...,\theta_q)$ are the generators of $\Lambda(q)$, we say that $(x_1,...,x_p,\theta_1,...,\theta_q)$ are the coordinates of $\mathcal{U}^{p,q}$.\\
A supermanifold, $\mathcal{M}$, is a ringed space $\left( M,\mathcal{O}_\mathcal{M} \right)$, where $\mathcal{O}_\mathcal{M}$ is a sheaf of commutative superalgebras and the following conditions are satisfied:\\
\begin{enumerate}
\item
$M$ is a Hausdorf space with a countable base.
\item
Every point $m$ in $M$ has a neighborhood U, such that the ringed space $\left( U,\mathcal{O}_\mathcal{M}|_U \right)$ is isomorphic to a superdomain $\mathcal{U}$.
\end{enumerate}
A smooth function on a supermanifold can be written as
\be f=\sum_v{f_v(x)\theta^v} \label{superfunction}\ee
Where $x={x_1,...,x_p}$, $v={v_1,...v_q}$, $v_i\in{0,1}$, $\theta^v=\theta_1^{v_1}\cdot...\cdot\theta_q^{v_q}$ and the sum is over all possible values of $v$. An integral of a function $f$ on a supermanifold $\mathcal{M}$ is defined as \be \int_\mathcal{{M}}f = \int_M f_{1,1,...,1}\label{superint} \ee
A derivative of a function of even and odd coordinates is defined as:
\be \partial_{x_i}f=\sum_v{\partial_{x_i}(f_v(x))\theta^v}; \partial_{\theta_i}f=\sum_v{v_i(-1)^{\sum_{j=1}^{i-1}}(f_v(x))\cdot\theta_1\cdot...\cdot\theta_{i-1}\cdot\theta_{i+1}\cdot...\cdot\theta^q} \ee
Note that  a derivative with respect to the even variables are even (commuting), whereas that with respect to the grassmanian variables is odd. With this the tangent space at a point $m$ of (the underlying manifold $M$ of) a supermanifold $\mathcal{M}$ is the space spanned by the derivatives at $m$. As shown above, it is a super vector space of dimension $(p,q)$. It should be noted that a more rigorous definition of the tangent space exists, but it is very technical, and the definition used here suffices. Having defined the tangent space, the tangent bundle, $T\mathcal{M}$ is defined in the usual manner. It is a $(2p,2q)$ dimensional supermanifold.\\
More important to us than the tangent space is the cotangent space. It is a space derived from the tangent space by flipping the parity of all the generators. We denote the generators of the cotangent space by $dx_1,...,dx_p,d\theta_1,...,d\theta_q$. Note again that now the $d\theta$'s are even, commuting variables, whereas the $dx$'s are grassmanian. The cotangent bundle, denoted by $\Pi TM$ is, in a manner similar to the purely even case, a $(p+q,p+q)$ dimensional manifold.\\
A pseudodifferential form on a supermanifold $\mathcal{M}$ is a function on $\Pi TM$. In a fashion akin to (\ref{superfunction}), such a function can be written as
\be f=\sum_{v,u}{f_{v,u}(x,d\theta)\theta^v dx^u} \label{pseudodiff} \ee
Integration of a pseudodifferential form is defined just like in (\ref{superint}), and integrating only the variables along the cotangent space is called integration along a fibre, or fibrewise integration. Notice that the $dx$'s are grassmanian and therefore pose no problem for the integration. The $d\theta$'s, however, are even variables, and the fibre is a linear space, and so for the integral along the fibre to converge we need $f_{1,1,....,1}$ to be rapidly decreasing in those variables, i.e. for fixed $x$ that function decreases to zero faster than any polynomial in the $d\theta$'s. In our analysis we use fibrewise integration when doing a "super Fourier-Mukai" transformation, and this condition would be satisfied.

\section{Solution of the defect equations of motion in the fermionic case}
In section 4  we obtained the defect  equations of motion for the defect implementing fermionic T-duality:
\be\label{ftd1c}
E_{j1}\partial X^j-(-)^{s_j}E_{1j}\bar{\partial} X^j-\partial_{\tau}\tilde{\theta}^1=0
\ee

\be\label{ftd2c}
E_{jN}\partial X^j-(-)^{s_js_N}E_{Nj}\bar{\partial} X^j-\tilde{E}_{jN}\partial \tilde{X}^j+(-)^{s_js_N}\tilde{E}_{Nj}\bar{\partial} \tilde{X}^j=0
,\hspace{0.5cm} N=2\ldots {\rm dim} M
\ee

\be\label{ftd3c}
\tilde{E}_{j1}\partial \tilde{X}^j-(-)^{s_j}\tilde{E}_{1j}\bar{\partial} \tilde{X}^j+\partial_{\tau}\theta^1=0
\ee

\be\label{fdxtc}
\partial X^N+\bar{\partial} X^N=\partial \tilde{X}^N+\bar{\partial} \tilde{X}^N,\hspace{0.5cm} N=2\ldots {\rm dim} M
\ee
Writing separately  terms with  $\theta^1$ in (\ref{ftd1c}) and with $\tilde{\theta^1}$  in (\ref{ftd3c}) and using (\ref{ftdual}) we get
\be\label{ftd12c}
E_{M1}\partial X^M-(-)^{s_M}E_{1M}\bar{\partial} X^M+B_{11}(\partial\theta^1+\bar{\partial}\theta^1)-(\partial\tilde{\theta}^1+\bar{\partial}\tilde{\theta}^1)=0
\ee
\be\label{ftd22c}
E_{M1}\partial \tilde{X}^M-(-)^{s_M}E_{1M}\bar{\partial} \tilde{X}^M-(\partial\tilde{\theta}^1+\bar{\partial}\tilde{\theta}^1)+B_{11}(\partial\theta^1+\bar{\partial}\theta^1)=0
\ee
Taking sum and difference of  (\ref{ftd12c}) and (\ref{ftd22c}) we obtain:
\be
(E_{M1}+(-)^{s_M}E_{1M})(\bar{\partial} \tilde{X}^M-\bar{\partial} X^M)=0
\ee
and
\be
E_{M1}\partial \tilde{X}^M+B_{11}\partial\theta^1-\partial\tilde{\theta}^1-(\bar{\partial}\tilde{\theta}^1-B_{11}\bar{\partial}\theta^1+(-)^{s_M}E_{1M}\bar{\partial} \tilde{X}^M)=0
\ee
Separating terms with  $\theta^1$ and $\tilde{\theta^1}$ also in (\ref{ftd2c}) we obtain
\bea
&& {E_{1N}\over B_{11}}\left(E_{M1}\partial \tilde{X}^M+B_{11}\partial\theta^1-\partial\tilde{\theta}^1\right) \\ \nonumber
&& +{E_{N1}(-)^{s_N}\over B_{11}}\left(\bar{\partial}\tilde{\theta}^1-B_{11}\bar{\partial}\theta^1+(-)^{s_M}E_{1M}\bar{\partial} \tilde{X}^M\right)+ \left(E_{MN}+(-)^{s_Ms_N}E_{NM}\right)\left(\bar{\partial} \tilde{X}^M-\bar{\partial} X^M\right)=0
\eea
Collecting all we get:
\bea
&&\bar{\partial} \tilde{X}^N=\bar{\partial} X^N,\hspace{0.5cm} N=2\ldots {\rm dim} M\\ \nonumber
&&\partial \tilde{X}^N=\partial X^N,\hspace{0.5cm} N=2\ldots {\rm dim} M\\ \nonumber
&&\partial\tilde{\theta}^1=B_{11}\partial\theta^1+E_{M1}\partial X^M\\ \nonumber
&&\bar{\partial}\tilde{\theta}^1=B_{11}\bar{\partial}\theta^1-(-)^{s_M}E_{1M}\bar{\partial} X^M
\eea

\end{document}